\documentclass{PoS}

\usepackage{url}
\def\arxiv#1{\href{http://arxiv.org/abs/#1}{{\tt arXiv:#1}}}

\title{Target of Opportunity Observations of Blazars with H.E.S.S.}

\ShortTitle{Blazar ToO observations with H.E.S.S.}

\author{\speaker{F. Sch\" ussler$^1$} and M. Seglar-Arroyo$^1$\\
             $^1$ IRFU, CEA,
Universit\'e Paris-Saclay,
F-91191 Gif-sur-Yvette, France\\
     E-mail:  \email{fabian.schussler@cea.fr},
     \email{monica.seglar-arroyo@cea.fr}}

\author{M. Arrieta,$^2$, M. Boettcher$^3$, C. Boisson$^2$,M. Cerruti$^4$, N. Chakraborty$^5$,  I.D. Davids$^3 ,^6$, F. Jankowsky$^7$, J.P. Lenain $^4$, H. Prokoph $^8$, D. Sanchez $^9$, S. Wagner $^{7}$, M. Zacharias $^{3}$ and A. Zech$^2$ on behalf of the H.E.S.S. Collaboration\\
E-mail: \email{contact.hess@hess-experiment.eu}\\
\\
\llap{$^2$}LUTH, Observatoire de Paris, PSL Research University, CNRS, Universit\'e Paris Diderot, 5 Place Jules Janssen, 92190 Meudon, France \\
\llap{$^3$}Centre for Space Research, North-West University, Potchefstroom 2520, South Africa\\
\llap{$^4$}Sorbonne Universit\'es, UPMC Universit\'e Paris 06, Universit\'e Paris Diderot, Sorbonne Paris Cit\'e, CNRS, Laboratoire de Physique Nucl\'eaire et de Hautes Energies (LPNHE), 4 place Jussieu, F-75252, Paris Cedex 5, France\\
\llap{$^5$}Max-Planck-Institut f\"ur Kernphysik, P.O. Box 103980, D 69029 Heidelberg, German\\
\llap{$^6$}University of Namibia, Department of Physics, Private Bag 13301, Windhoek, Namibia\\
\llap{$^7$}Landessternwarte, Universit\"at Heidelberg, K\"onigstuhl, D 69117 Heidelberg, Germany\\
\llap{$^8$}GRAPPA, Anton Pannekoek Institute for Astronomy and Institute of High-Energy Physics, University of Amsterdam,  Science Park 904, 1098 XH Amsterdam, The Netherlands\\
\llap{$^9$} Laboratoire d'Annecy-le-Vieux de Physique des Particules (LAPP), Univ. de Savoie,
CNRS/IN2P3, Annecy-le-Vieux F-74941, France}

\abstract{The very-high-energy (VHE, E>100 GeV) extragalactic sky is dominated by blazars, a class of active galactic nuclei which show rapid variability at all wavelengths. Target of Opportunity (ToO) observations triggered by flaring activity detected at longer wavelengths are, thus, an important part of the blazar observing strategy of H.E.S.S., an array of five imaging atmospheric Cherenkov telescopes sensitive to VHE photons. \\
In this contribution we detail the H.E.S.S. extragalactic ToO program, describing the specific procedures currently in place to follow up on multi-wavelength alerts. The program is illustrated by discussing a few recent noteworthy targets observed with the H.E.S.S. phase II array over the last two years of blazar ToO observations.}

\FullConference{35th International Cosmic Ray Conference --- ICRC2017\\
		10--20 July, 2017\\
		Bexco, Busan, Korea}

\begin{document}

\section{Introduction}
The extragalactic sky observed at very high energies (VHE) with Imaging Air Cherenkov Telescopes (IACTs) like H.E.S.S. is dominated by blazars, a sub-class of active galactic nuclei (AGN). They are phenomenologically characterized by non-thermal emission extending from radio to $\gamma$-rays, showing extreme variability both in frequency and timescale and polarized emission in optical wavelengths. These features are explained by assuming that these objects are radio-loud AGN hosting a jet of plasma that is expelled in relativistic speeds and whose axis points towards the Earth. This results in a boosted photon emission over all other AGN components, i.e. accretion disk, broad-line regions (BLR) or X-ray corona.\\

Blazars are usually divided into two different classes, according to the presence or absence of emission lines from the BLR in the optical spectrum and thus, to their relative strength with respect to the non-thermal emission. BL Lacertae objects, building the majority of the blazars detected at VHE, do not present these emission lines, while Fast Spectrum Radio Quasars (FSRQ) do. They are typically more powerful and their radiative output tend to be dominated by high-energy emission.\\

The spectral energy distributions (SED) of blazars typically extend over the entire electromagnetic spectrum, from radio to $\gamma$-ray energies, and present a characteristic broad double bump shape. The low energy component peaks, depending on the source, between the IR and the UV/soft X-ray band and it is typically assumed to be synchrotron emission from electrons and positrons in the jet. The high-energy component peaks between MeV and TeV energies, reaching its maximum in the $\gamma$-band, and it is thought to be produced either by inverse Compton mechanisms from the same leptons producing the low-energy component (cf. leptonic models \cite{leptonic}) or from accelerated high-energy hadrons loosing energy through synchrotron emission or photo-meson reactions (cf. \cite{hadronic1,hadronic2}). While FSRQs typically present a low frequency synchrotron peak in the infrared range, BL Lacertae present a variety of peak frequencies allowing to a classification in low/intermediate/high-frequency-peaked objects (LBL/IBL/HBL). \\

The emission of blazars shows strong variability over more than 20 orders of magnitude and in all frequency ranges from radio to $\gamma$-rays. They rank therefore among the most variable objects in the universe. Indeed, although blazar variability time scales reach from years down to minutes, the most violent outbursts last typically for a few days. This short time-scale variability allows to probe several interesting aspects of these sources like particle acceleration, cosmic-ray interactions and radiation mechanisms. Over the last years the H.E.S.S. collaboration put in place a dedicated Target of Opportunity (ToO) program to allow for rapid reaction and dedicated gamma-ray follow-up of triggers from observations across the electromagnetic spectrum. We here outline this program and report latest results focusing on observations obtained in 2016.\\

During the 2016 season, the H.E.S.S. experiment dedicated $\sim$60 hours of observation on ToOs of flaring blazars, which represents $\sim$11\% of the time allocated to extragalactic observations. The program incorporates a large variety of information that is monitored and used to trigger VHE observations on flaring blazars. It can be summarized as follows:

\begin{itemize}
\item Public alerts coming from the MWL community, distributed for example via Astronomer's Telegrams and/or Gamma-ray Coordinates Network (GCN).
\item Public data from several experiments like Swift-XRT and FACT. 
\item Optical data obtained with the dedicated ATOM telescope located at the H.E.S.S. site.
\item Private alerts from MWL partners, namely MAGIC, VERITAS, FACT, HAWC shared in the context of Memoranda of Understanding (MoU).
\item Automatic, daily analysis of \textit{Fermi}-LAT data using aperture photometry and a full likelihood analysis. 
\end{itemize}

In this work, a few highlights from H.E.S.S. ToO observation during the 2016 season are presented: the detection of PKS 1749+096 by H.E.S.S., the detection of enhanced gamma-ray emission of PKS 0447-439, first discovered by H.E.S.S in December 2009~\cite{PKS0447detection} and upper limits on the VHE emission of the FSRQs PKS 2022-07 and CTA 102.

\section{The H.E.S.S experiment}
H.E.S.S. is a stereoscopic system consisting of five IACTs located at 1800 meters above sea level in the Khomas highlands of Namibia. Starting operations in 2004, the initial configuration consisted in four 12 m telescopes arranged in a square of 120 m side length. In its second phase a fifth, 28 m telescope was added to the center of the array in 2012. In its initial four-telecope configuration, H.E.S.S. was sensitive to gamma-ray energies from 100 GeV to about 100 TeV while the energy threshold of the H.E.S.S. II phase is $\sim$ 30 GeV. Being the first hybrid Cherenkov instruments, it has the ability of taking data in different modes and configurations.\\

The general IACT detection technique is based on the detection of VHE $\gamma$-rays by studying the Cherenkov cascade of charged particles that are created when the photon interacts with particles present in the Earth's atmosphere. In order to discriminate between hadrons-triggered and photon-triggered showers, width, length and other characteristics of the showers are used. Details of the analyses of the showers and the discrimination technique can be found for the \textit{Model} analysis used in the presented work in \cite{Model} and for the independent analysis chain used for cross-checks, \textit{ImPACT}, in \cite{Impact}.\\

The analysis of the H.E.S.S. II data varies depending on the configuration. In the monoscopic mode used throughout the analyses presented here, only information from the fifth, 28 m telescope is used, while in stereoscopic mode the full array data is considered. The low-energy threshold provided by the monoscopic mode is particularly beneficial for studies of distant objects like AGNs, their flaring activity and, concretely, for the study of FSRQ and LBLs, characterized by a soft VHE spectrum. 

\section{PKS 1749+096: the July 2016 flare}

PKS1749+096, also known as OT081, is a bright radio-source, classified as low frequency-peaked BL Lacertae object, that shows weak extended jet emission to the northeast of the compact VLBI core on parsec scales \cite{PKS1749}. Redshift measurements from \cite{PKS1749z} give a value of z = 0.32. It shows radio emission typical of blazars: a core-jet morphology and superluminal motion of the jet components detected being in the range of 5-21c \cite{PKS1749}.\\

\begin{figure}[!h]
\vspace{-1mm}
\begin{minipage}{0.46\linewidth}
\centerline{\includegraphics[width=0.92\linewidth]{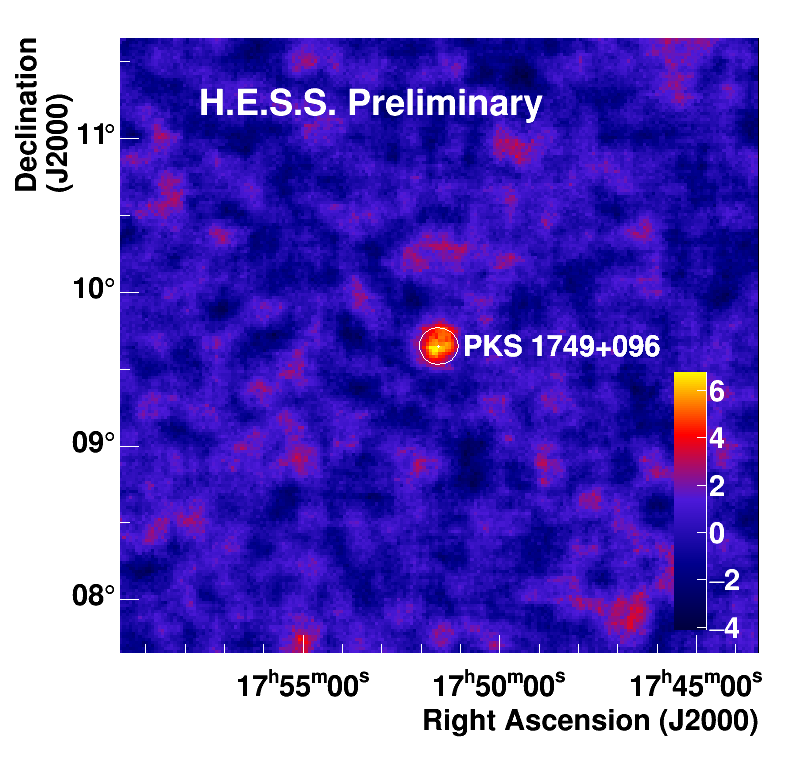}}
\end{minipage}
\hfill
\begin{minipage}{0.54\linewidth}
\centerline{\includegraphics[width=1.\linewidth,height=0.7\linewidth]{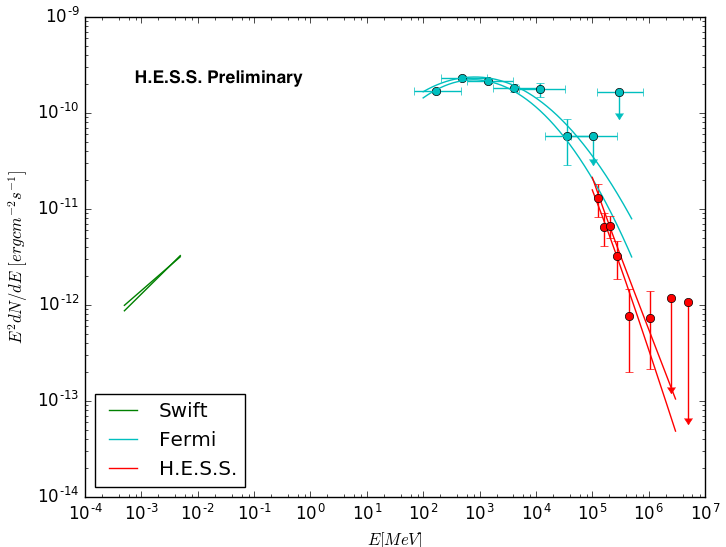}}
\end{minipage}
\caption{\footnotesize{Left: Significance map of PKS 1749+096 obtained from H.E.S.S. observations in July 2016. Right: Multi-wavelength spectral energy distribution of PKS 1749+096 derived from contemporaneous Swift-XRT, Fermi-LAT and H.E.S.S. observations. }
\label{fig:PKS1749MapSED} }    
\end{figure}

PKS 1749+096 started to show bright flaring activity on July 9th 2016 in the MeV-GeV energy band with Fermi-LAT as reported in ATel \# 9231. Following this unprecedented gamma ray activity, observations in X-ray and UV were carried on the 17th and 20th of July as described in ATel \#9259 and the time correlated gamma-ray/X-ray/UV/optical outburst confirmed the source indentification of  3FGL J1751.5+0939 with the BL Lac object OT 081. At that time, due to moon constraints, it was not visible for H.E.S.S.. Observations could start on July 22th and H.E.S.S. monitored the source for 6 consecutive nights. We continued to monitor the source until July 27th and obtained a total of $\sim$ 12 hours of observations.  As reported in ATel \#9267, MAGIC detected VHE emission from PKS 1749+096 during the night of July 24th. Here we report H.E.S.S. observations of the same period, showing a clear detection of the source above the $5\sigma$ level for the full observation period. The Li\&Ma significance map derived from the observations is shown in the left plot of Fig.~\ref{fig:PKS1749MapSED}. We can also confirm the detection of a flaring flux increase during the night of 24th of July, where we reach a detection above $6\sigma$ in only 2.3 h of observations. It is visible in the VHE light-curve shown in the upper panel of Fig.~\ref{fig:PKS1749_LC}.

\subsection{Multi-wavelength context}

The H.E.S.S. ToO program does not only consist in triggering VHE observations but the responsible team is also in charge of assuring a broad multi-wavelength coverage. Various means are used for this: optical observation in various bands with the onsite ATOM telescope, analysis of available data from other observatories like Fermi-LAT, proposal submission to instruments like Swift-XRT, etc. 

\begin{figure}[!t]
\centering
\includegraphics[width=0.95\textwidth,height=0.75\textwidth]{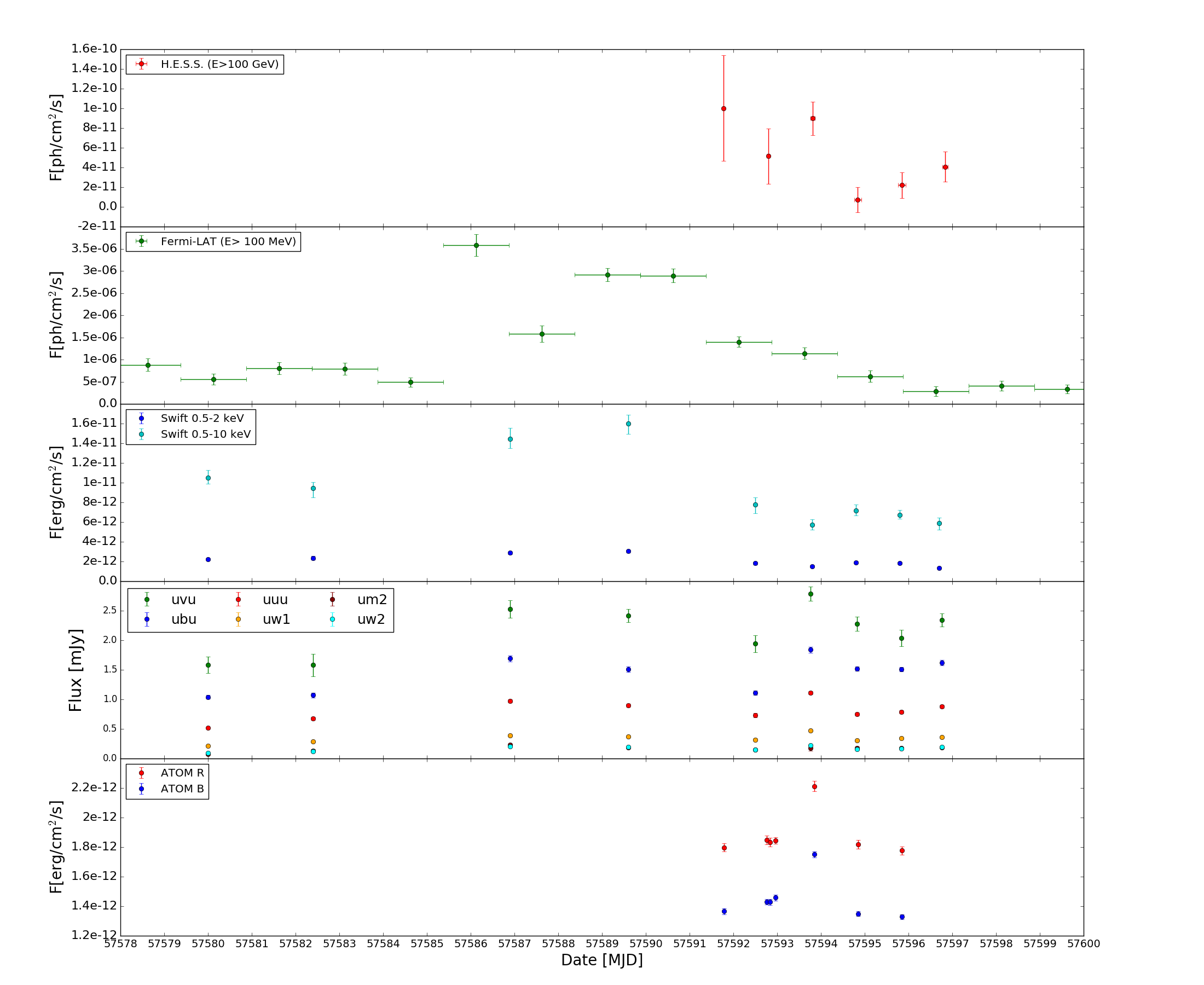}
\caption{\footnotesize{Multi-wavelength light-curves of PKS1749+096 in July 2016, from top to bottom in decreasing energy ranges: VHE observations with H.E.S.S., HE data obtained with Fermi-LAT, dedicated X-ray observations obtained with Swift-XRT and Swift-UVOT and optical observations from ATOM. }}
\label{fig:PKS1749_LC}     
\end{figure}

We show here the multi-wavelength data obtained during and around the H.E.S.S. observation campaign of PKS 1749+096 as illustration. The spectral energy distribution obtained from Swift-XRT, Fermi-LAT, derived by using the Enrico tool \cite{SanchezEnrico}, and H.E.S.S. observations is shown in the right plot of Fig.~\ref{fig:PKS1749MapSED}. The obtained light-curves are summarized in Fig.~\ref{fig:PKS1749_LC}, which shows from top to bottom observations in decreasing energy ranges: VHE observations with H.E.S.S., HE data obtained with Fermi-LAT, dedicated X-ray observations obtained with Swift-XRT and Swift-UVOT and optical observations from ATOM. The flare seen in the VHE band seems to be only very weakly correlated with activity in other bands, except the optical, where a coincident flux increase is found. Further, detailed analysis of the behavior of PKS 1749+096 during its flare in July 2016 is in preparation.

\section{PKS 0447-439: the October 2016 flare}
PKS 0447+096 is a very bright HE gamma-ray source detected by Fermi-LAT and presents a hard spectrum in the HE band following a power-law behavior with an spectral index of $2.77 \pm 0.28$~\cite{FermiLAT_PKS0447}. This BL-Lac object was detected in VHE gamma-rays in 2009 by H.E.S.S. \cite{PKS0447detection}. At that time the redshift of the object was not known and we used the gamma-ray data to derive constrains. Combining H.E.S.S. and Fermi-LAT data and considering the attenuation caused by gamma-ray interactions with the extragalactic background light (EBL) an upper limit of $z > 0.59$ could be derived~\cite{PKS0447_EBL}. A value of $z=0.343$ has later been proposed based on correlations with nearby galaxies studied with the Gemini Multi-Object Spectrograph~\cite{GMOS}.\\

\begin{figure}[!t]
\vspace{-1mm}
\begin{minipage}{0.5\linewidth}
\includegraphics[width=1.\textwidth,height=0.68\textwidth]{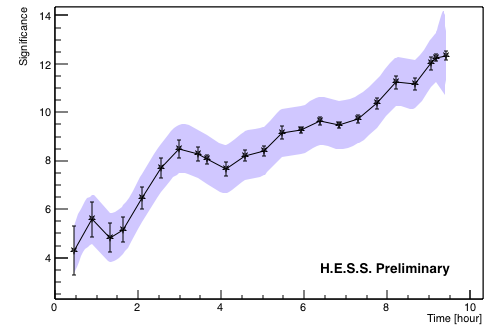}
\end{minipage}
\hfill
\begin{minipage}{0.5\linewidth}
\includegraphics[width=0.95\textwidth,height=0.68\textwidth]{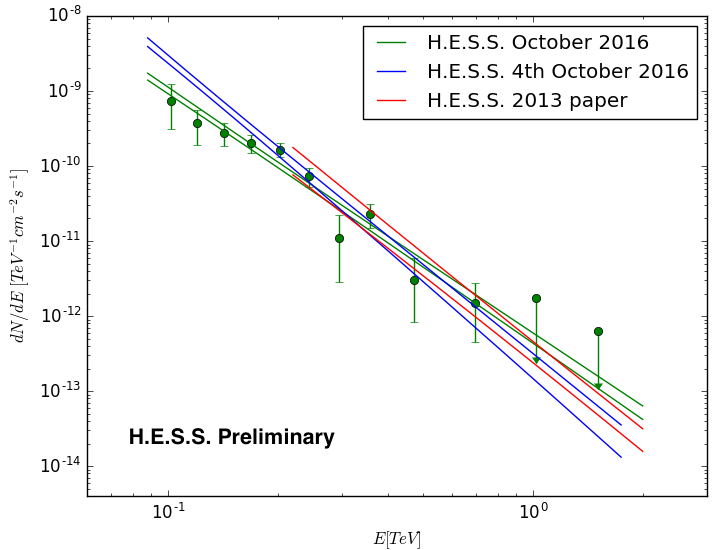}
\end{minipage}
\hfill
\caption{\footnotesize{Left: Significance evolution with time of PKS 0447-439 showing a total 12.3 sigma detection of the source and a bump on  October 4th. Right: Gamma-ray spectrum of PKS 0447-439 obtained in October (green markers) agrees well with archival observations with H.E.S.S. in 2009  (red lines). Variability on the gamma-ray spectrum on the 4th is presented (blue lines)}}
\label{fig:PKS0447MapSED}      
\end{figure}

In 2016, PKS 0447-439 was part of a monitoring program of several extragalactic objects. During these monitoring observations a significant flaring outburst was observed on October 4th using a rapid, on-site analysis of the obtained data. A H.E.S.S. internal self-trigger was issued and the source was observed over several days between October 4th and October 9th. A total of 9.4h of data could be recorded. The evolution of the significance of the gamma-ray excess with time is shown in the left plot of Fig.~\ref{fig:PKS0447MapSED}. The bump visible for the night of the October 4th indicates a flaring state. PKS 0447-439 was clearly detected at a significance level of $12\,\sigma$ above an energy threshold (defined as $10\%$ of the maximum acceptance) of about $80\,\mathrm{GeV}$. The derived energy spectrum for the dataset without the night of the flare is well described by a power-law with photon index $\Gamma$ =  -3.29 $\pm$ 0.18 and a differential flux at 1 TeV equal to $(5.19 \pm 1.62) \cdot 10^{-13}\,\mathrm{cm}^{-2}\,\mathrm{s}^{-1}\,\mathrm{TeV}^{-1}$ with statistical uncertainties only. As illustrated in the right plot of Fig.~\ref{fig:PKS0447MapSED}, the spectrum agrees well with the data obtained at the end of 2009~\cite{PKS0447_EBL}. During the night of the flare (October 4th, 2016), the energy spectrum is best described by a slightly softer power-law with photon index $\Gamma$ =  -4.06 $\pm$ 0.39, leading to a slightly lower differential flux at 1 TeV ($(2.34 \pm 1.89) \cdot 10^{-13}\,\mathrm{cm}^{-2}\,\mathrm{s}^{-1}\,\mathrm{TeV}^{-1}$).


\section{PKS 2022-077: the April 2016 flare}
PKS 2022-077 is a FSRQ at redshift $z=1.388$~\cite{PKS2022z}. In April 2016 a HESS ToO campaign was launched on this source, following reports on high flux state in gamma-rays by AGILE (ATeL \#8879, March 28) and Fermi LAT (ATeL \#8932, April 10) and in the radio domain by RATAN-600 (ATeL \#8893, April 1). The GeV flare, according to Fermi-LAT observations, started around April 9th and lasted for about 11 days. During this period the GeV flux reached approximately 5 times the one from the Crab nebula in this energy band.

\begin{table}[!ht]
    \caption{\footnotesize{Upper limits on the VHE gamma-ray flux derived from H.E.S.S. observations of PKS 2022-077 in April 2016.}}    \label{table1}
\begin{center}
    \begin{tabular}{ | c | c | c |}
    \hline
    E [GeV] & ULs in E$^2$ dN/dE [erg/cm$^2$/s] & EBL-deabsorbed ULs in E$^2$ dN/dE [erg/cm$^2$/s]    \\ \hline
    129 & 1.2 $\cdot$ 10$^{-11}$ & 8.5  $\cdot$ 10$^{-11}$ \\ \hline
    156 & 1.7  $\cdot$ 10$^{-11}$ & 25.7   $\cdot$ 10$^{-11}$ \\ \hline
    188 & 2.9  $\cdot$ 10$^{-11}$ & 117.9 $\cdot$ 10$^{-11}$  \\ \hline
    227 &2.6 $\cdot$ 10$^{-11}$&341.4 $\cdot$ 10$^{-11}$\\ 
    \hline
    \end{tabular}
\end{center}
\label{default}
\end{table}%

The source was visible to HESS at a favorable zenith angle only starting around April 12th. Note that, due to the very high redshift of this source, EBL absorption effects are significant and strong zenith angle constrains ($\theta < 40\,\mathrm{deg}$) were imposed in order to achieve an as low as possible energy threshold. A total of $1'5\,\mathrm{h}$ of observations with an energy threshold of about $E> 110\,\mathrm{GeV}$ could be obtained during the nights of April 13th and 14th, 2016. No significant VHE gamma-ray emission has been detected. The resulting differential upper limits on the flux can be found in Table \ref{table1}. The EBL model of~\cite{EBLmodel} has been used to correct the observed limits for absorption effects.

\section{CTA 102: the August 2016 flare}
\begin{figure}[!ht]
\begin{minipage}{0.49\linewidth}
\includegraphics[width=1.0\textwidth,height=0.79\textwidth]{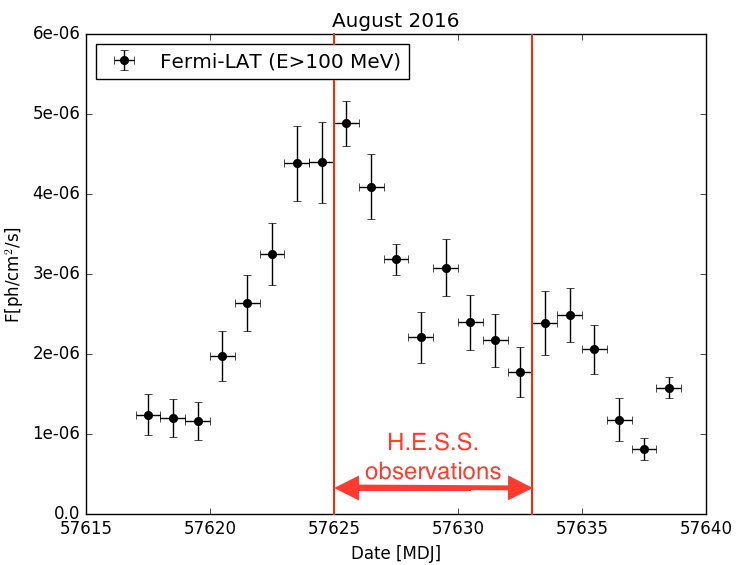}
\end{minipage}
\hfill
\begin{minipage}{0.49\linewidth}
\vspace{2mm}
\includegraphics[width=0.9\textwidth,height=0.79\textwidth]{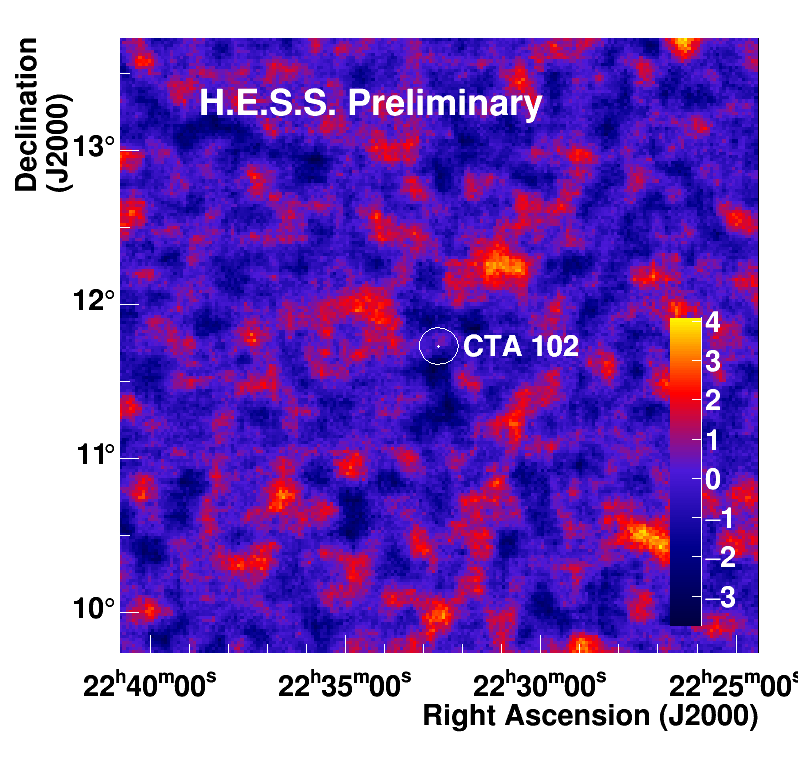}
\end{minipage}
\caption{\footnotesize{Left: Fermi-LAT light-curve above 100 MeV of CTA 102 during the August 2016 flare. Right: Significance map showing no detection of CTA 102 in the VHE range observed by H.E.S.S.}}
\label{fig:CTA102_GeVLC_Map}     
\end{figure}

CTA 102 is a Flat-Spectrum Radio Quasar at a redshift z= 1.037~\cite{CTA102z}. An optical outburst has been detected in August 2016 by ATOM at the H.E.S.S. site. The flare was also seen in GeV gamma-rays by the Fermi-LAT instrument, and the light-curve above $100\,\mathrm{MeV}$ of CTA 102 in August 2016 is illustrated in the left plot of Fig.~\ref{fig:CTA102_GeVLC_Map}. 

H.E.S.S. ToO observations of CTA 102 started in August 25th and lasted for about one week. The obtained $10\,\mathrm{h}$ of data didn't reveal significant VHE emission above an energy threshold of about $100 \,\mathrm{GeV}$. The derived significance map is shown in the right plot of Fig.~\ref{fig:CTA102_GeVLC_Map}. 

\section{Acknowledgements}
{\footnotesize The support of the Namibian authorities and of the University of Namibia in facilitating the construction and operation of H.E.S.S. is gratefully acknowledged, as is the support by the German Ministry for Education and Research (BMBF), the Max Planck Society, the German Research Foundation (DFG), the Alexander von Humboldt Foundation, the Deutsche Forschungsgemeinschaft, the French Ministry for Research, the CNRS-IN2P3 and the Astroparticle Interdisciplinary Programme of the CNRS, the U.K. Science and Technology Facilities Council (STFC), the IPNP of the Charles University, the Czech Science Foundation, the Polish National Science Centre, the South African Department of Science and Technology and National Research Foundation, the University of Namibia, the National Commission on Research, Science \& Technology of Namibia (NCRST), the Innsbruck University, the Austrian Science Fund (FWF), and the Austrian Federal Ministry for Science, Research and Economy, the University of Adelaide and the Australian Research Council, the Japan Society for the Promotion of Science and by the University of Amsterdam.
We appreciate the excellent work of the technical support staff in Berlin, Durham, Hamburg, Heidelberg, Palaiseau, Paris, Saclay, and in Namibia in the construction and operation of the equipment. This work benefited from services provided by the H.E.S.S. Virtual Organisation, supported by the national resource providers of the EGI Federation.
}

\end{document}